\documentclass[]{article}
\usepackage{emulateapj}
\usepackage{graphicx}
\usepackage{multirow}

\advance \voffset by -0.5cm\relax

\usepackage{color}
\def\msun{{\rm ~M}_{\odot}}

\def\zsun{{\rm ~Z}_{\odot}}
\def\zsunn{{\rm Z}_{\odot}}

\begin{document}

\title{Comparison of LIGO/Virgo upper limits with predicted compact binary
  merger rates}

 \author{Krzysztof Belczynski\altaffilmark{1,2},
         Michal Dominik\altaffilmark{1}, 
         Serena Repetto\altaffilmark{3},
         Daniel E. Holz\altaffilmark{4,5},
         Christopher Fryer\altaffilmark{5}}

 \affil{
     $^{1}$ Astronomical Observatory, University of Warsaw, Al.
            Ujazdowskie 4, 00-478 Warsaw, Poland\\
     $^{2}$ Center for Gravitational Wave Astronomy, University of Texas at
            Brownsville, Brownsville, TX 78520, USA\\
     $^{3}$ Department of Astrophysics/IMAPP, Radboud University Nijmegen, PO Box 9010, 
            6500 GL Nijmegen, The Netherlands\\
     $^{4}$ Enrico Fermi Institute, Department of Physics, and Kavli Institute
            for Cosmological Physics, University of Chicago, Chicago, IL 60637\\
     $^{5}$ Theoretical Division, Los Alamos National Laboratory,
            Los Alamos, NM 87545\\
 }

\begin{abstract}
We compare the current LIGO/Virgo upper limits on double 
compact object volumetric merger rates with our theoretical 
predictions. Our optimistic models are a factor of $\sim 3$ below the 
existing upper limits for massive BH-BH systems with total mass $50$--$70\msun$,
suggesting that a small increase in observational sensitivity may bring the first 
detections.
The LIGO/Virgo gravitational wave observatories are currently being upgraded to
advanced design sensitivity. 
If a sizeable population of BH-BH binaries is detected, the maximum 
total binary mass of this population will discriminate between two general families of 
common envelope models. If no binaries are detected, the new upper limits will provide 
astrophysically useful information about the environment and physical processes (e.g., 
metallicity of host galaxies or BH natal kicks) crucial to the formation of binaries 
containing black holes. For NS-NS systems, our predicted rates are $\sim 3$ orders of 
magnitude below the current upper limits; even if advanced instruments reach 
their design sensitivities (factors of $10\times$ in distance, and 
$1,000\times$ in volumetric rate) the detection of NS-NS systems is not assured.
However, we note that although our predicted NS-NS merger rates are consistent with 
estimates derived from Galactic NS-NS binaries and short GRBs, they are on the low 
side of these empirical estimates.
\end{abstract}


\section{Introduction}

Most massive stars are found in binary systems (Goodwin et al. 2007). During the
evolution of these stars the binaries can experience component merger during common 
envelope (CE) phases (Webbink 1984) or disruption during supernova (SN) 
explosions (Tauris \& Takens 1998) in which individual stars form neutron stars
(NSs) or black holes (BHs). The massive binaries which survive these processes 
form double compact objects: NS-NS, BH-BH, or mixed BH-NS systems. These remnant 
systems are subsequently subject to angular momentum loss via the emission of 
gravitational waves (GWs) and their orbital separation decreases (Peters \& Mathews 
1963; Weisberg \& Taylor 2005). Finally, the two compact objects collide and merge 
into a single compact object giving rise to a strong GW signal (Einstein 1918). 
The LIGO/Virgo network of ground-based interferometric observatories has been 
designed to search for these signals.\footnote{http://www.ligo.caltech.edu/; 
https://wwwcascina.virgo.infn.it/}

Initial LIGO/Virgo observations were concluded in $2012$ without the detection of a 
GW signal. The instruments are currently being upgraded and the network will resume 
its operation in several years with advanced sensitivity. Various predictions for 
near-future detection chances were compiled and presented by Abadie et al. (2010). 

One of the most promising sources for these advanced GW detectors is the inspiral 
and merger of binary NS systems. There are several known Galactic double neutron 
star binaries with merger times shorter than the Hubble time (e.g., Kim, Kalogera 
\& Lorimer 2010). Double black hole binaries (BH-BH), on the other hand, remain 
undetected. Recent theoretical predictions indicate that these systems may either 
dominate GW observations (e.g., Belczynski et al. 2010a) or be totally absent in 
local Universe (e.g., Mennekens \& Vanbeveren 2013). 

In this study we compare the rates from our evolutionary calculations of double 
compact object mergers  (Dominik et al. 2012) with the existing LIGO/Virgo 
upper limits (Abadie et al. 2012; Aasi et al. 2013). We then make predictions
for the science that will be probed with future upper limits and/or detections 
of double compact object mergers with advanced GW instruments.

\section{Calculations}
 
We have employed a set of publicly available evolutionary models
({\tt www.syntheticuniverse.org}) that provide physical properties and Galactic 
merger rates (${\cal R}_{\rm MW}$) of NS-NS, BH-NS and BH-BH binaries. The 
calculations of the merger rates were obtained with the {\tt StarTrack} 
population synthesis  code (Belczynski et al. 2002, 2008), with the inclusion of 
crucial updates in the physical models (winds, SN, CE). We have chosen $12$ 
physically realistic models, each testing one unknown in the evolutionary process 
leading to the formation of a NS-NS/BH-BH binary. For each model $2 \times 10^6$ 
primordial binaries were evolved, and we have converted the resulting Galactic 
merger rates to the volumetric local Universe merger rates using the formula: 
\begin{equation}
{\cal R}_{\rm vol} = 10^{-6} \rho_{\rm gal} \left( f_{\rm Z} 
{\cal R}_{\rm MW}^{\zsunn} + (1-f_{\rm Z}) {\cal R}_{\rm MW}^{0.1\zsunn} \right)
{\rm Mpc}^{-3} {\rm yr}^{-1}
\end{equation} 
where ${\cal R}_{\rm MW}^{\zsunn}$ is the Galactic merger rate for models with 
solar-like metallicity ($Z=0.02$) and ${\cal R}_{\rm MW}^{0.1\zsunn}$ is the 
Galactic merger rate for models with low metallicity ($Z=0.002$). The Galactic merger 
rates (${\cal R}_{\rm MW}^{\zsunn}, {\cal R}_{\rm MW}^{0.1\zsunn}$) are available 
from the aforementioned database and are expressed in Myr$^{-1}$. 
The fraction of the local stellar population with solar metallicity, as opposed to 
$0.1\zsunn$, is given by $f_{\rm Z}$. By altering this quantity we can construct 
local stellar populations with  a range of metal content. We assume that a local 
density of Milky Way-like galaxies is $\rho_{\rm gal}=0.01$ Mpc$^{-3}$. 

The projected volumetric merger rates are listed in Table~\ref{rates}, sorted into 
bins of total binary mass ($M_{\rm tot}$). The top row lists the LIGO upper limits 
on the rates from Abadie et al. (2012; see top panel of their Fig.4) for 
$M_{\rm tot}<25\msun$, and from Aasi et al. (2013; see left panel of their Fig.5) 
for larger masses. These upper limits are for equal mass binaries; they are therefore 
the most stringent for a given total binary mass.

The standard model, denoted model S, employs the current best estimates for various 
physical parameters, including some that are not yet fully constrained but play an 
important role in the formation of double compact objects. For example: during CE 
evolution physical estimates of the donor binding energy are used (Xu \& Li 2010), 
we adopt $M_{\rm NS,max}=2.5 \msun$ as the maximum NS mass (Lattimer \& Prakash 2010), 
NS natal kicks are taken from observations as a single Maxwellian with $\sigma=265$ 
km s$^{-1}$ (Hobbs et al. 2005), BH kicks are smaller and obtained through a mass 
ejection mechanism (Fryer et al. 2012), the compact object mass spectrum is based on 
rapid supernova explosions (Belczynski et al. 2012), the stellar winds are revised for 
the effects of clumping (Belczynski et al. 2010b), and mass transfer episodes are 
non-conservative with $50\%$ of the mass retained in the binaries (Meurs \& van den 
Heuvel 1989). 

In comparison to the standard model, we also run the following models. 
In models V5 and V6 we adopt $M_{\rm NS,max}=3.0, 2.0 \msun$, respectively. 
Natal kicks are decreased to $\sigma=132.5$ km s$^{-1}$ in V7 for NSs and BHs. 
High BH kicks ($\sigma=265$ km s$^{-1}$) and no BH kicks are adopted in V8 and V9, 
respectively. Delayed supernovae are employed in V10. The wind mass loss rate is
decreased by factor $2$ in V11. Conservative and fully non-conservative mass
transfer is assumed in V12 and V13, respectively. 

The investigation of $\sim 30,000$ Sloan Digital Sky Survey galaxies revealed that 
recent ($\sim 1$ Gyr) star formation was bimodal with about half stars formed with 
high and half with low metallicity (Panter et al. 2008). Therefore, in all the
aforementioned models we have evolved a $50\%$--$50\%$ combination of two stellar
populations, one with high and one with low metallicity ($f_{\rm Z}=0.5$; see eq.1).
To test the influence of metallicity on our predictions we also use a $90\%$--$10\%$ 
combination of metallicities, in which most stars are formed mostly at high 
metallicity ($f_{\rm Z}=0.9$). In model V16 this high metallicity combination is 
applied to the standard model while in model V17 it is applied to a high BH natal 
kick model (V8). 

For each model we have tested  whether the CE phase prevents the formation of 
NS-NS/BH-NS/BH-BH binaries (Belczynski et al. 2007). In particular, in the A models
we allow for binary survival in the CE phase even if a donor was a Hertzsprung gap (HG) 
star. It is not fully understood whether such stars have clear core-envelope structure
and whether they will behave like MS stars (always resulting in a CE merger) or evolved 
giant stars (with potential CE survival). In the B models we assume that an HG donor 
always leads to a CE merger.

\section{Results}

All of our models are fully described in Dominik et al. (2012). In what follows
we list the most noteworthy trends, as well as compare directly with existing
observational upper limits.

Our standard model calculation is presented in Figure~\ref{std}. The dependence of the merger 
rate density on the total mass of the binary begins with a pronounced peak in the first 
mass bin ($M_{\rm tot}=2$--$5\msun$) where all of the NS-NS systems and some of the 
BH-NS systems are found. The next mass bin ($M_{\rm tot}=5$--$8\msun$) with relatively 
low merger rate densities contains mostly BH-NS systems. Then for higher mass bins 
merger rate density increases and the dependence is relatively flat, with a sharp rate 
density drop off at $M_{\rm tot} \sim 70\msun$ and $\sim 50 \msun$ for models A and B, 
respectively. Rate densities in model B are a factor of a few smaller than in model A. 
Model A rate densities are high, with ${\cal R}_{\rm vol} \sim 10^{-8}$--$10^{-7}$  
Mpc$^{-3}$ yr$^{-1}$. In particular, the predicted merger rate densities are only a 
factor of $\sim 3$ below the upper limits from LIGO/Virgo in the $M_{\rm tot}=54$--$73\msun$ 
bin. Note that the predicted rates in the lowest mass bin are $\sim 3$ and $\sim 2.5$ 
orders of magnitude below the upper limits for models A and B, respectively. 

A number of evolutionary models closely resemble the standard model: V5, V6, V7, V9, 
and V10 (see Table~\ref{rates} for comparison). Several other models differ in various 
details from the standard model, but show an overall agreement in merger rate densities: 
V11, V12, and V13. Below we describe those models with significant differences from the 
standard model. These models represent a sequence of decreasing rate densities for 
massive binaries with $M_{\rm tot}>10\msun$. 

In model V14 we have altered the metallicity distribution of stellar populations in the 
local Universe to one with $90\%$ of stars at high metallicity and $10\%$ of stars at low 
metallicity. A majority of BH-BH binaries in our models form from low metallicity stars 
due to the lower wind mass loss rates and increased chance of CE survival (Belczynski et al. 
2010a; Dominik et al. 2012). In model V14, therefore, we note an order of magnitude decrease
(from the standard model, S) in the rate densities for massive mergers ($M_{\rm tot}>10\msun$), 
resulting in rates of $\sim 10^{-8}$ Mpc$^{-3}$ yr$^{-1}$ (see Fig~\ref{metals}). 

In model V8 we have increased the magnitude of BH natal kicks. Since this is one
of the key factors in BH-BH formation, the results in a significant decrease in
the rate density, to $\sim 10^{-9}-10^{-8}$ Mpc$^{-3}$ yr$^{-1}$ (see Fig.~\ref{kicks}). 
A significant increase in natal kicks (factor of $\sim 4$; see Discussion) leads to 
the disruption of progenitor binaries, thereby inhibiting BH-BH formation. In this 
model we do not use the kick mechanism based on asymmetric mass ejection during the 
supernova explosion that was adopted in our other models. During NS formation, 
exploding stars tend to be massive ($10$--$20\msun$) in relation to the NS mass 
($1.4 \msun$), and even a small asymmetry in the large ejected mass imparts a 
significant momentum kick on a proto-NS. At BH formation the picture is qualitatively 
different, as the mass of a dying star is similar to the mass of the BH that is being 
formed. Most of the star mass is removed prior to the  BH formation via intense 
stellar winds; e.g., Luminous Blue Variable phase and Wolf-Rayet phase (these winds 
are not expected for NS progenitors). Therefore very little mass ejection is predicted 
at BH formation, and natal kicks are expected to be insignificant. On the other hand, 
significant neutrino emission may be expected at formation of both neutron stars and 
black holes, possibly leading to high natal kicks in both cases. 

Finally, in model V15 we have combined these two factors, high metallicity and high BH 
kicks, that limit BH-BH formation. As expected, this leads to a severe decrease in the 
rate density of massive mergers, to $\sim 10^{-10}-10^{-9}$ Mpc$^{-3}$ yr$^{-1}$ (see 
Fig.~\ref{kicks2}. In particular, the rate densities are so low that even if the current 
LIGO/Virgo upper limits are increased by factor of $1,000$ (corresponding to a factor of 
$10$ in instrument distance sensitivity), no BH-BH mergers would be predicted to be 
detected.

\section{Discussion}

\subsection{NS-NS merger rates}

Our predicted rates for NS-NS mergers are presented in Figures~\ref{std}, \ref{metals}, 
\ref{kicks}, \ref{kicks2} and the first mass bin of Table~\ref{rates}. These rates, 
although low, are consistent with available 
empirical estimates. Kim et al. (2010) have estimated NS-NS merger rate based on 
observations of three Galactic field NS-NS systems, B1913+16, B1534+12, and J0737-3039, 
and found a Galactic  merger rate within the range $3$--$190$ Myr$^{-1}$. O'Shaughnessy 
\& Kim (2010) obtained a median value of $89$ Myr$^{-1}$, with a spread above and below 
by a factor of $\sim 3$ when pulsar beaming constraints are taken into account. Kim, 
Perera \& McLaughlin (2013) have re-examined the influence of double pulsar J0737-3039, 
obtaining a revised estimate of the Galactic merger rate of $7$--$49$ Myr$^{-1}$, with 
median value $21$ Myr$^{-1}$. None of these estimates include the large uncertainties in 
the pulsar luminosity function. If these uncertainties are included, it is expected that 
the rates could shift up or down by an order of magnitude (Richard O'Shaughnessy 2013, 
private communication). Applying this to the most recent estimate results in a broad
range of allowed rates: $2.1$--$210$ Myr$^{-1}$. For comparison, the Galactic merger 
rates in our standard evolutionary scenario are $23.5$ and $7.6$ Myr$^{-1}$ for model A 
and B, respectively. Note that we list here only the rates for solar metallicity ($Z=0.02$; 
i.e., Table 2 of Dominik et al. 2012) as these are relevant for Galactic field evolution 
where NS-NS binaries are found. Our NS-NS merger rates for sub-solar metallicity 
($Z=0.002$; i.e., Table 3 of Dominik et al. 2012) are factor of $\sim 3$ lower. The full 
spread of our rates for the solar metallicity models used in this study is $23.3$--$77.4$ 
Myr$^{-1}$ for the A models and $0.3$--$9.5$ Myr$^{-1}$ for the B models. Our predicted 
merger rates, while consistent with Galactic NS-NS observations, are on the lower end of 
empirical estimates.

There is a mounting evidence that short Gamma-ray bursts (GRBs) are connected with NS-NS 
and/or BH-NS mergers 
(e.g., Berger 2013), with some authors using short GRB rates to estimate NS-NS merger rates 
(e.g., Chen \& Holz 2013). Fong et al. (2013) found short GRB rate densities at the level 
$100$--$1,000$ Gpc$^{-3}$ yr$^{-1}$, while Petrillo, Dietz, \& Cavagila (2013) estimated 
the rate density to be $500$--$1,500$ Mpc$^{-3}$ yr$^{-1}$. These results suffer short GRB
beaming and luminosity uncertainties. The beaming has been firmly established for $\sim 3$ 
short GRBs (with $\theta_j\lesssim 10\,\deg$), while redshifts and thus luminosities are 
only known for the $\sim 20$ closest events. It is possible that the average beaming angle 
is larger than currently estimated, and that the rates densities are correspondingly lower, 
possibly by an order of magnitude (Edo Berger 2013, private communication). The lower limit 
on the short GRB rate density would then decrease to $10$--$50$ Gpc$^{-3}$ yr$^{-1}$. Before 
comparing with our NS-NS merger rate density it is worth noting that even if NS-NS stars are 
in fact short GRB progenitors, they may be responsible for only a fraction of short GRBs
as other progenitors cannot be excluded at the moment (e.g., Nakar 2010). On the other hand 
only some fraction of NS-NS mergers may produce short GRBs. The NS-NS merger rate densities 
that we have adopted from Dominik et al. (2012) are at the level of $50$ and $150$ Gpc$^{-3}$ 
yr$^{-1}$ for our standard model calculation for submodels B and A, respectively. We have 
employed a $50\%$--$50\%$ combination of high and low metallicity stellar populations 
($f_{\rm Z}=0.5$) since short GRBs are found in host galaxies with a range of metallicities. 
Note that the NS-NS rate density can only be approximately read off Table~\ref{rates}, as the 
first mass bin contains both NS-NS and BH-NS mergers. However, NS-NS mergers dominate in 
this bin, so the first mass bin is a good approximation for the overall NS-NS merger rate 
density. The clear cut division between NS-NS and BH-NS merger rates is given in Dominik et 
al. (2012). Again, we note that our rate predictions are consistent with short GRB 
observations, although they are found somewhat on the low side of empirical estimates.

\subsection{BH natal kicks}

The current models behind natal kicks involve asymmetric mass ejection in supernova 
explosions leading to high NS kicks and small BH kicks (e.g., Herant, Benz, Colgate 1992; 
Blondin, Mezzacappa, DeMarino 2003). However, at the moment the possibility that BHs 
receive high natal kicks through some other mechanism (e.g., asymmetric neutrino emission; 
Lai 2001 and references within) cannot be excluded on theoretical grounds.

In our standard model we employ natal kicks that are associated with asymmetric mass 
ejection, while in model V8 we assume high BH natal kicks (e.g., from asymmetric neutrino 
emission). Comparison of rate densities for both models clearly demonstrates a significant 
decrease of BH-BH systems due to high BH kicks. We conclude that natal BH kicks are one of 
the most important factors in determining the number of BH-BH systems.

Thus far the only empirical approach to get insights into the natal kicks black holes 
receive at birth, is achieved by following both the Galactic dynamics and the binary 
evolution of black-hole X-ray binaries (BH-XRBs), which are semi-detached binaries where a 
companion star, either of low-mass or high-mass, is transferring material to a black hole. 
The natal kick at birth changes the Galactic position and the space velocity of the binary,
as well as the binary orbital properties. Tracing the binary evolution and the trajectory
of the binary backwards in time, it is then possible to place constraints on the natal kick 
the black hole receives. This approach has been applied in the past to a handful of sources,
and seems to point towards a dichotomous scenario for the black-hole natal kicks. Small 
natal kicks, of the order of few tens of km s$^{-1}$ are found for GRO J1655-40 (Willems et 
al. 2005), and for Cygnus X-1 (Wong et al. 2012); whereas evidence for an intermediate to high 
natal-kick is found for XTE J1118+480, a BH-XRB which requires a natal kick in the range 
$80-310$ km s$^{-1}$ (Fragos et al. 2009). Repetto, Davies \& Sigurdsson (2012) performed an 
analysis of positions and velocities of most known Galactic  BH-XRBs. They constrain the natal 
kick by making sure that the binary after-kick orbital configuration is consistent with the 
observed one, and at the same time that the natal kick is high enough to place the binary in 
the current height from the Galactic plane, assuming it was born on the plane. The natal kick 
they obtain is an absolute lower-limit, the space velocity in the post-supernova configuration 
being in the optimal direction, i.e. perpendicular to the Galactic plane. They found that $9$ 
BHs do not require any or only a small kick ($\sim 0-40$ km s$^{-1}$; XTE J1550-564, GRO 
J1655-40, GRS 1915+105, GS 2023+338, GRO J0422+32, A0620-003, GRS 1009-45, 1124-683, GS 
2000+251), $2$ BHs require intermediate kicks ($\sim 80$ km s$^{-1}$; 4U 1543-47, XTE J1118+480) 
and the other two require high kicks ($190, 450$ km s$^{-1}$ for 1819.3-2525 and 1705-250, 
respectively) comparable to pulsar kicks.
Repetto \& Nelemans (in preparation) combines the study of BH-XRB positions and velocities 
with detailed binary evolution calculation. They trace the binary evolution of the systems 
backwards until the post-supernova configuration. Their results are consistent with the ones 
in Repetto et al. (2012), and show evidence for black hole receiving intermediate or high 
kicks in at least  some of the sources.
In Table~\ref{T:kicks} we show the collected data on BH natal kick estimates along with BH 
mass estimates for 14 Galactic BH-XRBs. The natal kick estimates are lower limits on the 
natal kick, the space velocity being in the optimal direction.

In Figure~\ref{bhkick} we show these observational estimates along with our standard and 
high BH kick models. Both the BH kicks and the mass estimates suffer large uncertainties 
which we choose not to plot to allow a clearer comparison of trends in the observations and 
theoretical predictions. For the theoretical models we show the average values of the kick. 
High BH kicks are modeled as a (mass independent) 1D Maxwellian distribution of kicks with 
$\sigma=265$ km s$^{-1}$, with an average at $420$ km s$^{-1}$. For our standard model the 
kick is drawn from a Maxwellian with $\sigma=265$ km s$^{-1}$, with the value decreased 
inversely proportional to the amount of fall back mass at BH formation in the core 
collapse/supernova explosion. The amount of fall back is estimated from rapid supernova 
models that reproduce the observed mass gap between NSs and BHs in Galactic X-ray binaries 
(Belczynski et al. 2012). The non-monotonic dependence of fall back on BH (and progenitor) 
mass is discussed and explained in Appendix. As clearly seen from Figure~\ref{bhkick}, while 
neither of the models can explain all the empirical estimates, the standard model with low 
BH kicks associated with asymmetric mass ejection appears to be a better match to the 
available observations.

\section{Conclusions}

\begin{figure}
\includegraphics[width=0.7\columnwidth]{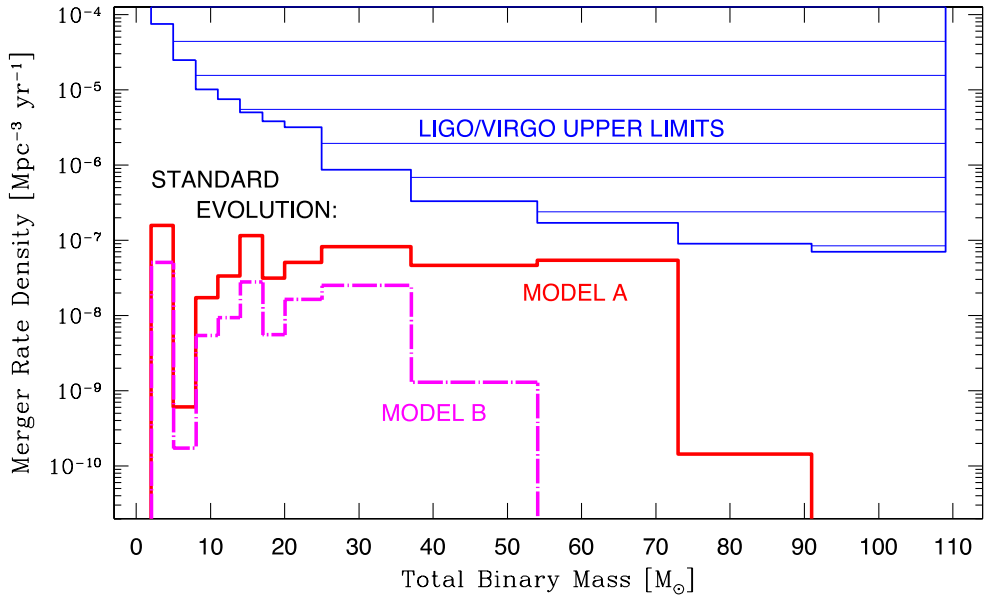}
\vspace*{0cm}
\caption{
Rate density of double compact object mergers for our standard evolutionary
calculation. The solid line represents our population synthesis model in which we 
allow for CE survival with HG donors (model A). The dashed-dotted line shows 
a model in which CE events with HG donors are not allowed (model B). Most of 
our evolutionary variations (e.g., V5, V6, V7, V9, V10) are indistinguishable 
from the standard prediction. Note that model B is characterized by lower rate
densities (down by a factor of a few in each mass bin) and a significant rate
density drop off occurs at much smaller total binary mass ($M_{\rm tot}=35\msun$) 
than for model A ($M_{\rm tot}=70\msun$). The top shaded area shows the upper limits
from initial 
LIGO/Virgo. 
}
\label{std}
\end{figure}

Our results show that a small improvement in LIGO/Virgo's current sensitivity to
double compact object mergers, by a factor of $\sqrt[3]{3}$ in distance (factor
of 3 in volume), will
either bring the first detections of massive BH-BH binaries ($M_{\rm tot}>50\msun$) 
or through non-detection exclude 
entire families of evolutionary models (see the proximity of Model A to the
LIGO/Virgo upper limits in Figure~\ref{std}). 

\begin{figure}
\includegraphics[width=0.7\columnwidth]{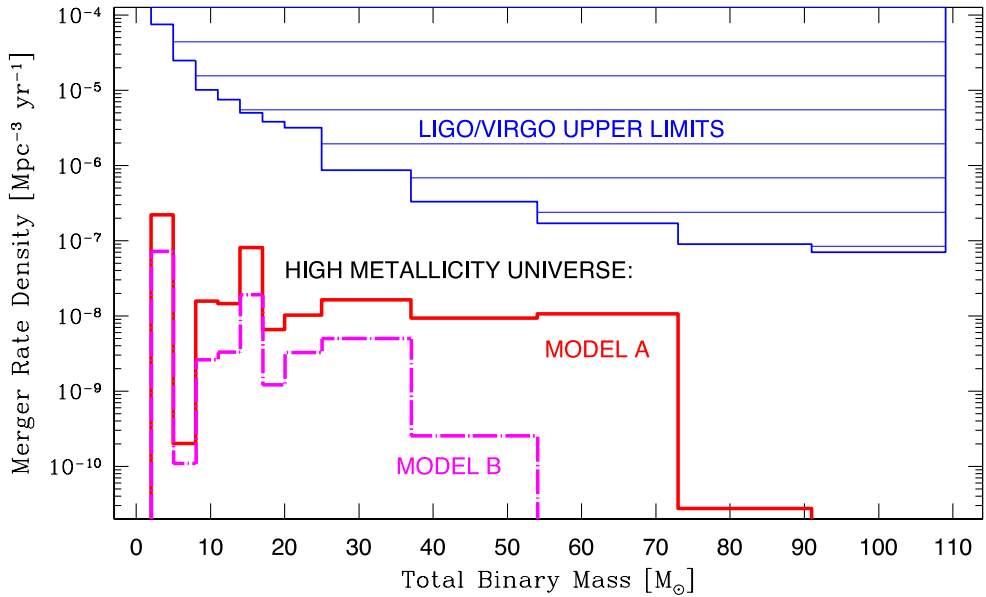}
\vspace*{0cm}
\caption{
Rate density of double compact object mergers for our standard evolution with
one change---we consider a toy universe with a high content of high metallicity stars: 
$90\%$ of stars with $Z=\zsun$ and $10\%$ with $Z=0.1\zsun$ (V14). For comparison,
the rest of our models were obtained with a low content of high metallicity
stars: $50\%$ of stars with $Z=\zsun$ and $50\%$ with $Z=0.1\zsun$.
Note the moderate decrease in rate density of high mass mergers in both the A and 
B models as compared with the low metallicity standard model (Fig.~\ref{std}). 
} 
\label{metals}
\end{figure}

\begin{figure}
\includegraphics[width=0.7\columnwidth]{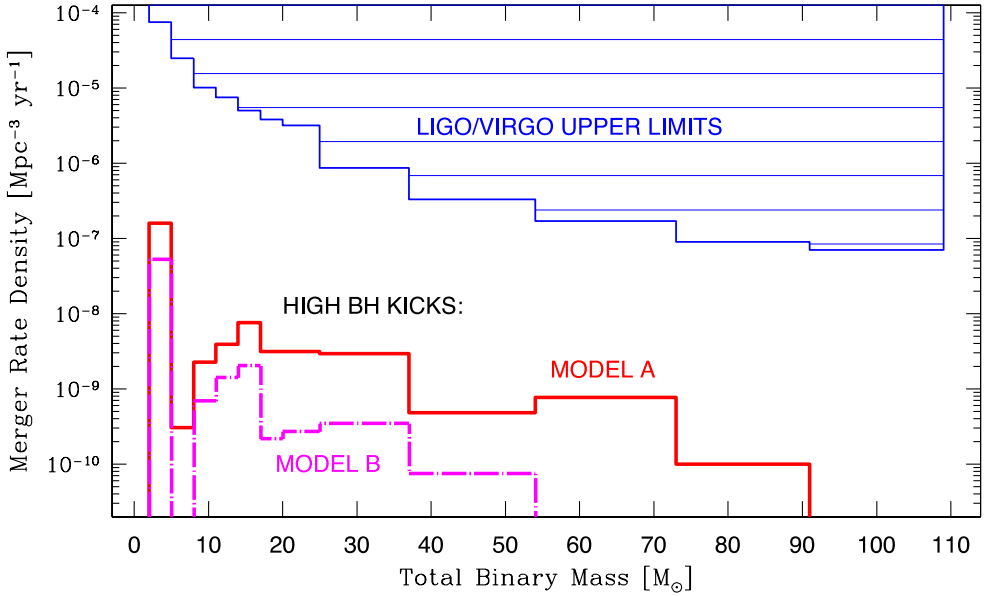}
\vspace*{0cm}
\caption{
Rate density of double compact object mergers for the high BH kick model (V8). 
Note the significant decrease in rate density for both the A and B models as compared 
with the standard evolutionary scenario (low or no BH kicks). However, the distinctive
behavior of models A and B (i.e. differing maximum mass of merging binary)
is the same as in the standard model case. 
} 
\label{kicks}
\end{figure}

If discoveries of massive BH-BH binaries are made, even at low rates, it will 
indicate that stars just beyond the main sequence (i.e., on the Hertzsprung gap) survive common 
envelope and form close double compact objects (see Figures~\ref{std}, \ref{metals}, 
\ref{kicks}, and differences between models A and B). 

If upper limits are improved by a factor of $100$ and we remain without the
detection of BH-BH binaries, it will indicate that other factors than common envelope are responsible for 
eliminating the BH-BH formation channels. Our analysis indicates that high black hole natal 
kicks and high metallicity stellar environments are two crucial factors that significantly 
reduce the BH-BH merger rates (see Figures~\ref{metals}, \ref{kicks} and the associated 
merger rate reductions). 
These two factors show a degeneracy when it comes to the rate of BH-BH mergers,
so additional constraints are required to distinguish their relative importance. 

\begin{figure}
\includegraphics[width=0.7\columnwidth]{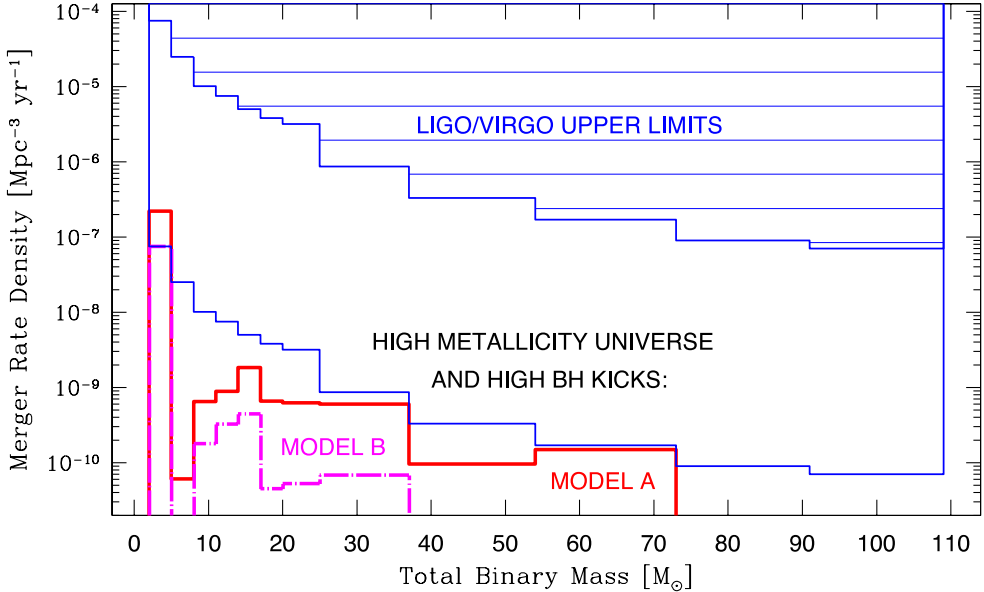}
\vspace*{0cm}
\caption{
Rate density of double compact object mergers for the combined high metallicity and 
high BH kick model (V15). We also show improved (by a factor of $1,000$) upper
limits expected from advanced sensitivity instruments. Note that in this model all BH-BH
binaries are just below the improved upper limits; it would not predict any detections. 
} 
\label{kicks2}
\end{figure}

\begin{figure}
\includegraphics[width=0.7\columnwidth]{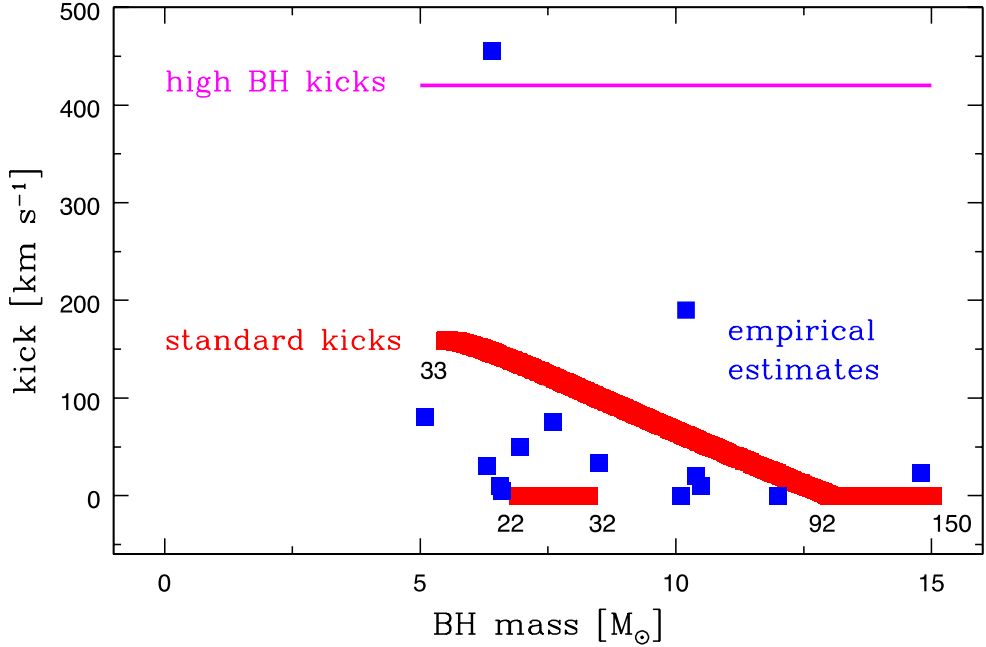}
\vspace*{0cm}
\caption{
Average black hole natal kick adopted in our calculations. In our standard model  
the kicks decrease with increasing mass of the black hole (red). Zero kicks (direct
collapse) are found for the highest mass black holes as well as for medium mass
black holes, the latter reflecting the physics of our core
collapse/supernova model (rapid explosion of Fryer et al. 2012). The kicks are
drawn from a Maxwellian distribution with a kick magnitude decreasing with 
increasing fall back mass estimated at each black hole's formation. 
In our high black hole kick model (magenta) the kicks are independent of the 
black hole mass and taken from a distribution with $\sigma=265$ km s$^{-1}$ 
(with an average kick of $420$ km s$^{-1}$).
Blue squares show available empirical estimates for Galactic black holes 
(see Table~\ref{T:kicks}; if two values of a kick are measured for an object then
an average is used in this plot). The red line representing our standard model 
shows kicks corresponding to single star solar metallicity evolution. Along this 
line we list initial (Zero Age Main Sequence) mass of the black hole progenitor in 
$\msun$ (see the Appendix for a detailed explanation).
} 
\label{bhkick}
\end{figure}

If upper limits are improved by factor of $1,000$ and no BH-BH 
discoveries are made, then the above degeneracy is lifted and we can conclude
that BHs receive high natal kicks {\em and}\/ that the stellar populations within 
reach of LIGO are typically of high (solar-like) metallicity (see
Figure~\ref{kicks2}, where the BH-BH 
rates fall below the enhanced upper limits). This is demonstrated by our V15 model in 
which we have increased the magnitude of BH natal kicks and the high metallicity content 
of the local Universe to the largest values with reasonable limits: it
is difficult to imagine BHs receiving higher kicks 
than NSs (see Figure~\ref{bhkick}), and less than $10\%$ of stars in the local Universe having
sub-solar metallicity (e.g., Panter et al. 2008). 

The above conclusions apply only within the framework of our evolutionary model. Our model 
successfully recovers the observed BH mass spectrum both in the Galaxy and in other environments 
(Belczynski et al. 2010b). It also provides a physical explanation for the existence of 
the mass gap between neutron stars and black holes, failing to produce compact
objects in the mass 
range $2$--$5\msun$ (Belczynski et al. 2012). However, our model is unable to reproduce 
the companion mass distribution observed in Galactic BH transients. The observed distribution 
peaks at $\sim 0.6\msun$ while most of our predicted distributions peak at $\sim 1\msun$. 
This discrepancy may arise from our poor understanding of magnetic braking in
these low mass companions along with uncertainties in low mass stellar  
models. It is also possible that the discrepancy arises from observational biases (Wiktorowicz, 
Belczynski, \& Maccarone 2013). These factors should not play a significant role in 
the modeling of massive stars and the formation of NS-NS, BH-NS, or BH-BH binaries.  

A different reduction factor in BH-BH formation was proposed by  Mennekens \& Vanbeveren 
(2013). These authors claim that strong Luminous Blue Variable (LBV) winds may inhibit
formation of close BH-BH binaries. They argue that the action of these very 
intense winds from BH progenitors leads to an increase in orbital separation
such that BH-BH 
binaries with merger times below a Hubble time no longer form. This finding is
hard to reconcile with observations, 
as a number of Galactic and extra-galactic binaries with black holes on close orbits are 
known (e.g., extra-galactic IC10 X-1 with $P_{\rm orbital}=33\,$h or Galactic GRO J0422+32 
with $P_{\rm orbital}=5.1\,$h both hosting massive $\gtrsim10\msun$ BHs).   
 
If NS-NS mergers are found before the advanced instruments reach their design sensitivity 
(a improvement by a factor of $\sim1,000$ in volume) it will indicate that our predicted 
NS-NS merger rates densities are too low. The astrophysical implications of such a 
potential finding are now under study (Dominik et al., in preparation). Alternatively, it is 
worth noting that even with an increase in the existing upper limits by factor of 
$\sim1,000$ some of our model predictions, as well as low values of rates
derived from observations of
Galactic NS-NS binaries and cosmic short GRBs, allow for the possibility of non-detection of NS-NS mergers 
(see Figure~\ref{kicks2}; some of our models, as well as empirical estimates,
allow for lower merger rate densities than those shown). 

In conclusion, there exist a wide range of possibilities for the rates of
  compact binary mergers in advanced detectors like LIGO 
and Virgo. The first detections may be made with small improvements over the 
instruments sensitivity. Alternatively, even at full advanced design sensitivity 
there may be no detections of double compact object mergers. Regardless of
whether there are detections or more stringent upper limits on the rates,
LIGO/Virgo will address a number of important science problems in our
understanding of compact binary formation and evolution.

\acknowledgements
KB and MD acknowledge support from NCN grant SONATA BIS and FNP professorial subsidy 
MASTER2013 and NASA Grant NNX09AV06A to the UTB. MD acknowledges support from NCN 
grant 2011/01/N/ST9/00383.
DEH acknowledges support from National Science Foundation CAREER grant PHY-1151836.
The work of C.F. was done under the auspices of the National Nuclear Security 
Administration of the U.S. Department of Energy, and supported by its contract 
DEAC52-06NA25396 at Los Alamos National Laboratory.
We woud like to thank T.Bulik, I.Mandel. T.Piran, P.Brady, A.Buonanno for critical 
and constructive comments on the results presented in this study.

\begin{deluxetable}{llllllllllllll}
\tablewidth{550pt}
\tablecaption{Double Compact Object Merger Rate Density [Mpc$^{-3}$ yr$^{-1}$] \tablenotemark{a}}
\tablehead{&&&&&&$M_{\rm tot}$&\hspace*{-0.5cm}$/\msun=$&&&&&\\ 
Model & Total & 2-5 \tablenotemark{b} & 5-8 & 8-11 & 11-14 & 14-17 & 17-20 & 20-25
& 25-37 & 37-54 & 54-73 & 73-91\tablenotemark{c}}
\startdata
LIGO \tablenotemark{d} & --- & 7.5e-5 & 2.5e-5  & 1.0e-5  & 7.5e-6  & 5.0e-6  & 3.8e-6  & 3.2e-6  & 8.7e-7  & 3.3e-7  & 1.7e-7  & 0.9e-7  \\ 
&&&&&&&&&&&&\\
S\hspace*{0.42cm}A  & 5.9e-7 & 1.6e-7 & 6.1e-10 & 1.7e-8  & 3.3e-8  & 1.2e-7  & 3.1e-8  & 5.1e-8  & 8.3e-8  & 4.7e-8  & 5.4e-8  & 1.4e-10 \\
\hspace*{0.50cm} B  & 1.4e-7 & 5.0e-8 & 1.6e-10 & 5.4e-9  & 9.3e-9  & 2.8e-8  & 5.4e-9  & 1.6e-8  & 2.5e-8  & 1.3e-9  & 0       & 0       \\
V5\hspace*{0.18cm}A & 5.9e-7 & 1.6e-7 & 7.1e-10 & 1.6e-8  & 3.3e-8  & 1.1e-7  & 3.0e-8  & 5.1e-8  & 8.5e-8  & 4.7e-8  & 5.5e-8  & 1.2e-10 \\
\hspace*{0.50cm} B  & 1.4e-7 & 5.2e-8 & 1.9e-10 & 5.2e-9  & 9.2e-9  & 2.9e-8  & 5.1e-9  & 1.6e-8  & 2.5e-8  & 1.5e-9  & 0       & 0       \\
V6\hspace*{0.18cm}A & 6.0e-7 & 1.6e-7 & 6.5e-10 & 1.7e-8  & 3.4e-8  & 1.2e-7  & 3.2e-8  & 5.5e-8  & 8.3e-8  & 4.5e-8  & 5.5e-8  & 1.9e-10 \\
\hspace*{0.50cm} B  & 1.4e-7 & 5.2e-8 & 1.7e-10 & 5.0e-9  & 8.5e-9  & 2.8e-8  & 5.2e-9  & 1.8e-8  & 2.4e-8  & 1.4e-9  & 0       & 0       \\
V7\hspace*{0.18cm}A & 7.1e-7 & 2.0e-7 & 1.5e-9  & 2.6e-8  & 5.2e-8  & 1.4e-7  & 4.6e-8  & 6.2e-8  & 8.4e-8  & 4.6e-8  & 5.3e-8  & 9.6e-11 \\
\hspace*{0.50cm} B  & 1.7e-7 & 5.8e-8 & 7.1e-10 & 1.1e-8  & 1.2e-8  & 3.0e-8  & 9.0e-9  & 2.0e-8  & 2.5e-8  & 1.4e-9  & 0       & 0       \\
V8\hspace*{0.18cm}A & 1.8e-7 & 1.6e-7 & 3.1e-10 & 2.3e-9  & 3.9e-9  & 7.6e-9  & 3.1e-9  & 3.1e-9  & 2.9e-9  & 4.8e-10 & 7.7e-10 & 1.0e-10 \\
\hspace*{0.50cm} B  & 5.7e-8 & 5.2e-8 & 0       & 6.9e-10 & 1.4e-9  & 2.0e-9  & 2.1e-10 & 2.7e-10 & 3.4e-10 & 7.4e-11 & 0       & 0       \\
V9\hspace*{0.18cm}A & 7.4e-7 & 1.6e-7 & 1.6e-9  & 2.3e-8  & 7.7e-8  & 1.6e-7  & 5.4e-8  & 6.8e-8  & 8.6e-8  & 4.7e-8  & 5.4e-8  & 1.2e-10 \\
\hspace*{0.50cm} B  & 1.9e-7 & 5.2e-8 & 1.1e-9  & 1.1e-8  & 1.9e-8  & 3.9e-8  & 1.6e-8  & 2.5e-8  & 2.6e-8  & 1.4e-9  & 0       & 0       \\
V10A                & 5.0e-7 & 1.7e-7 & 2.5e-9  & 1.1e-8  & 2.1e-8  & 2.3e-8  & 2.9e-8  & 5.2e-8  & 8.6e-8  & 4.8e-8  & 5.3e-8  & 1.5e-10 \\
\hspace*{0.50cm} B  & 1.3e-7 & 5.7e-8 & 1.3e-9  & 4.0e-9  & 7.6e-9  & 4.7e-9  & 7.7e-9  & 1.6e-8  & 2.6e-8  & 1.3e-9  & 0       & 0       \\
V11A                & 7.3e-7 & 1.6e-7 & 6.6e-10 & 1.2e-8  & 3.3e-8  & 1.1e-7  & 3.3e-8  & 7.5e-8  & 2.0e-7  & 5.0e-8  & 2.7e-8  & 3.6e-8  \\
\hspace*{0.50cm} B  & 1.6e-7 & 4.4e-8 & 0       & 2.3e-9  & 9.6e-9  & 3.0e-8  & 5.6e-9  & 1.8e-8  & 4.7e-8  & 4.2e-9  & 0       & 0       \\
V12A                & 8.8e-7 & 4.7e-7 & 1.4e-10 & 9.3e-9  & 2.2e-8  & 6.7e-8  & 4.5e-8  & 6.7e-8  & 8.7e-8  & 5.0e-8  & 6.3e-8  & 2.0e-10 \\
\hspace*{0.50cm} B  & 7.8e-8 & 2.3e-8 & 1.2e-10 & 6.5e-9  & 9.3e-9  & 1.4e-8  & 3.4e-9  & 7.7e-9  & 1.4e-8  & 8.8e-10 & 0       & 0       \\
V13A                & 6.0e-7 & 1.6e-7 & 4.7e-9  & 1.6e-7  & 4.9e-8  & 7.8e-8  & 3.5e-8  & 5.4e-8  & 4.4e-8  & 1.0e-8  & 6.3e-9  & 0       \\
\hspace*{0.50cm} B  & 3.6e-7 & 3.8e-8 & 2.1e-9  & 1.2e-7  & 2.8e-8  & 4.4e-8  & 2.4e-8  & 4.9e-8  & 4.3e-8  & 5.6e-9  & 0       & 0       \\
V16A                & 3.9e-7 & 2.2e-7 & 2.0e-10 & 1.6e-8  & 1.5e-8  & 8.1e-8  & 6.6e-9  & 1.0e-8  & 1.6e-8  & 9.3e-9  & 1.1e-8  & 2.7e-11 \\
\hspace*{0.50cm} B  & 1.1e-7 & 7.2e-8 & 1.1e-10 & 2.6e-9  & 3.2e-9  & 1.9e-8  & 1.2e-9  & 3.2e-9  & 5.0e-9  & 2.5e-10 & 0       & 0       \\
V17A                & 2.3e-7 & 2.2e-7 & 6.5e-11 & 6.4e-10 & 8.9e-10 & 1.8e-9  & 6.6e-10 & 6.2e-10 & 5.9e-10 & 1.0e-10 & 1.5e-10 & 1.7e-11 \\
\hspace*{0.50cm} B  & 7.5e-8 & 7.4e-8 & 0       & 1.7e-10 & 3.2e-10 & 4.4e-10 & 4.2e-11 & 5.2e-11 & 6.9e-11 & 1.2e-11 & 0       & 0       \\

\enddata
\label{rates}
\tablenotetext{a}{
Rate densities are given under the assumption that the CE phase initiated by Hertzsprung gap donors 
may lead to the formation of a double compact object binary (A) or always halts binary 
evolution (B).}
\tablenotetext{b}{
Our binning corresponds to the LIGO/Virgo low and high mass search bins.   
}
\tablenotetext{c}{
We do not show results for the last mass bin $M_{\rm tot}=91$--$109\msun$, as
the rates are zero in this bin for all models but V11A, the latter with a rate 
of $4.9 \times 10^{-11}$ Mpc$^{-3}$ yr$^{-1}$. For comparison, the LIGO/Virgo upper limit 
for this bin is $0.7 \times 10^{-7}$ Mpc$^{-3}$ yr$^{-1}$. 
}
\tablenotetext{d}{
Available initial LIGO/Virgo upper limits for equal mass mergers are listed.
}
\end{deluxetable}

\hspace*{.1cm}
\begin{deluxetable}{llll}
\tablewidth{350pt}
\tablecaption{Black Hole Mass and Natal Kick \tablenotemark{a}}
\tablehead{No & Name & Mass [$\msun$] & Natal Kick [km s$^{-1}$]}
\startdata
1)  & 4U 1543-47                 & 5.1  (1)    &      80 (2)    \\ 
2)  & GRO J1655-40               & 6.3  (3)    &    30 (4)  \\
3)  & H 1705-250 (Nova Oph 77)      & 6.4  (5)    & 450-460 (2,6) \tablenotemark{b} \\ 
4)  & GS 2000+251                & 6.55 (7)    &    0-20 (2,6)  \\
5)  & A0620-00 (V616 Mon)        & 6.6  (8)    &    0-10 (2,6)  \\
6)  & GRS 1124-68 (Nova Mus 91)    & 6.95 (9)    &   40-60 (2,6) \tablenotemark{c} \\
7)  & XTE J1118+480              & 7.6  (10)   &  70-80 (2,11) \\
8)  & GRS 1009-45                & 8.5  (12)   &   15-50 (2,6)  \\
9)  & GRS 1915+105               & 10.1 (13)   &       0 (2) \\
10) & XTE J1819-254 (V4641 Sgr)   & 10.2 (14)   &     190 (2) \tablenotemark{d} \\
11) & GRO J0422+32               & 10.4 (15)   &   10-30 (2,6) \\
12) & XTE J1550-564              & 10.5 (16)   &      10 (2) \\
13) & GS 2023+338 (V404 Cyg)     & 12   (17)   &    0 (2) \\
14) & Cyg X-1                    & 14.8 (18)   &    0-55 (19,20) \\  

\enddata
\label{T:kicks}
\tablenotetext{a}{
References for the mass and natal kick estimates are given in parentheses:
(1) Orosz et al. (1998), (2) Repetto et al. 2012, (3) Greene et al. (2001),   
(4) Willems et al. (2005), (5) Harlaftis et al. (1997), (6) Repetto \&
Nelemans (in prep), (7) Ioannou et al. (2004), (8) Cantrell et al. (2010), 
(9) Gelino et al. (2001), (10) Khargharia et al. (2013), (11) Fragos et al. 
(2009), (12) Macias et al. (2011), (13) Steeghs et al. (2013), (14) Martin 
et al. (2008), (15) Reynolds et al. (2007), (16) Li et al. (2013), (17) 
Shahbaz et al. (1994), (18) Orosz et al. 
(2011), (19) Mirabel \& Rodrigues (2003), (20) Wong et al. (2012).
}
\tablenotetext{b}{Marked as 1705-250 in Repetto et al. (2012).}
\tablenotetext{c}{Marked as 1124-683 in Repetto et al. (2012).}
\tablenotetext{d}{Marked as 1819.3-2525 in Repetto et al. (2012).}
\end{deluxetable}

\pagebreak

\pagebreak

\pagebreak

\pagebreak

\pagebreak

\clearpage
\appendix
\section{BH mass -- kick velocity relation}

In our standard model for natal kicks in core collapse SNe we employ a Maxwellian kick 
distribution with $\sigma=265\,\mbox{km s}^{-1}$, based on observed velocities of single Galactic 
pulsars (Hobbs et al., 2005). The mass of the remnant object may generate a gravitational potential
strong enough to prevent parts or all of the mass ejected during SN from reaching escape velocity. 
The matter falling back onto the remnant object will reduce the original (Maxwellian) kick velocity
due to conservation of momentum. The more massive the final (pre-SN) core of the star, the more 
fallback it generates. In the asymmetric mass ejection kick mechanism (adopted
here) this results in BH natal 
kicks being smaller than NS kicks. To account for this effect we use a simple
linear relation for the reduction of natal kick magnitude by the amount of fallback during a SN:
\begin{equation} \label{vkick}
V_k=V_{max}(1-f_{\rm fb}),
\end{equation}
where $V_k$ is the final magnitude of the natal kick, $V_{max}$ is
the velocity drawn from a Maxwellian kick distribution, and $f_{\rm fb}$
is the fallback factor. The values of $f_{\rm fb}$ range between 0--1, with 0
indicating no fallback/full kick and 1 representing total fallback/no kick (in this
case a direct collapse).

Fig.~\ref{bhkick} shows how BH natal kicks are related to the BH mass in a solar metallicity
environment for our standard model (red line). In this model the minimum mass of a BH
progenitor is $\sim 20\msun$ at Zero Age Main Sequence (ZAMS), and they produce
$\sim 7\msun$ remnants. At the end of their lives, stars of this mass develop extended, 
high density layers of oxygen and silicon. When the nuclear evolution is completed, the infall 
of these layers stalls the SN explosion, resulting in a direct collapse to a BH. No matter is 
ejected and therefore no natal kicks are present. As we increase progenitor mass toward higher 
values, the final silicon and iron core masses and sizes increase. This results in the 
production of more massive BH remnants also through direct collapse. This trend continues 
until the progenitor mass reaches $\sim 30\msun$ at ZAMS, corresponding to a BH mass of 
$\sim 8\msun$.

Above $30\msun$ this trend is temporarily disrupted. A star with such mass at ZAMS loses enough
mass in stellar winds to strip itself of its hydrogen envelope and expose its helium core. 
Such objects, Wolf-Rayet (WR) stars, continue to lose mass at very high rates
($\sim 10^{-5}\msun$ yr$^{-1}$), and this extensive mass loss may lower the binding energy of the 
star, making it easier to explode. 

Whereas mass-loss alters the star's fate from the outside in, instabilities in shell burning at the 
end of nuclear evolution can alter the fate of the star from within the core. As we increase  
progenitor mass the core of the star grows in size and forces the shell burning layers toward  
the surface. At $\sim 30\msun$ the core is large enough for the shell burning to occur at different 
temperatures. The extent and explosiveness of the shell burning is very sensitive to the exact 
conditions of the star such that shell burning can be very different for stars of nearly the same 
mass. The result is a dip in the final silicon and iron core masses. These smaller silicon and
iron cores can lead to stronger explosion energies and lower remnant masses. 
Our standard model includes both of these effects (wind mass loss and shell burning instabilities).
Their interplay results in the lowest amount of fallback during a SN explosion. As a consequence,
the resulting remnant BH has the lowest mass ($\sim 5\msun$) and the highest natal kick for 
a $\sim 30\msun$ progenitor. However,
note that both wind mass-loss and shell burning are open fields  
of research and their effects on remnant masses are likely to change quantitatively as we gain a 
more complete understanding of the underlying physics.

Increasing the progenitor mass beyond $30\msun$ gradually increases the amount of fallback matter,
which reduces the natal kicks and increases the final remnants mass. BH progenitors with masses 
over $\sim 90\msun$ at ZAMS produce final silicon and iron cores massive enough for full fallback 
(direct collapse). As a result the kick velocities decrease to zero. 
A more detailed description of the supernova modeling is given in Fryer et al. (2012).


\begin{references}

\reference{} Aasi, J., et al.\ 2013, Phys. Rev. D., 87, 022002
\reference{} Abadie, J., et al.\ 2010, CAQG, 27, 173001 
\reference{} Abadie, J., et al.\ 2012, Phys. Rev. D, 85, 082002
\reference{} Belczynski, K., Kalogera, V., Bulik, T.\ 2002, ApJ, 572, 407
\reference{} Belczynski, K., et al.\ 2007, ApJ, 662, 504
\reference{} Belczynski, K., et al.\ 2008, ApJS, 174, 223 
\reference{} Belczynski, K., et al.\ 2010a, ApJ, 715, L138
\reference{} Belczynski, K., et al.\ 2010b, ApJ, 714, 1217 
\reference{} Belczynski, K., Wiktorowicz, G., Fryer, C., Holz, D., \& Kalogera, 
             V.\ 2012, ApJ, 757, 91
\reference{} Berger, E.\ 2013, ARAA, 51, accepted (arXiv:1311.2603))
\reference{} Blondin, J., Mezzacappa, A., DeMarino, C.\ 2003, ApJ, 584, 971
\reference{} Cantrell, J., et al.\ 2010, ApJ, 710, 1127
\reference{} Chen, H.-Y., \& Holz, D.E.\ 2013, PRL, 111, 181101
\reference{} Dominik, M., et al.\ 2012, ApJ, submitted (arXiv:1202.4901)
\reference{} Einstein, A.\ 1918, Uber Gravitationswellen, Sitzungsberichte der 
             physikalisch-mathematischen Klasse, 1, 154
\reference{} Fong, W., et al.\ 2012, ApJ, 756, 189
\reference{} Fragos, T., Willems, B., Kalogera, V., Ivanova, N., Rockefeller, G., 
             Fryer, C.L.,\& Young, P.A. \ 2009, ApJ, 697, 1057
\reference{} Fryer, C., et al.\ 2012, ApJ, 749, 91 
\reference{} Gelino, D., Harrison, T., \& McNamara, B.\ 2001, ApJ, 122, 971
\reference{} Goodwin S. P., Kroupa P., Goodman A., Burkert A.\ 2007, 
             in Reipurth B., Jewitt D., Keil K., eds, Protostars and Planets V The
             Fragmentation of Cores and the Initial Binary Population, p.133
\reference{} Greene, J., Bailyn, C., \& Orosz, J.\ 2001, ApJ, 554, 1290
\reference{} Harlaftis, E., et al.\ 1997, AJ, 114, 1170
\reference{} Herant, M., Benz, W., \& Colgate, S.\ 1992, ApJ, 395, 642

\reference{} Hobbs, G., Lorimer, D. R., Lyne, A. G., \& Kramer, M.\ 2005, 
             MNRAS, 360, 974 
\reference{} Ioannou, Z., Robinson, E., Welsh, F., \& Haswell, C.\ 2004, AJ, 127, 481
\reference{} Khargharia, J., Froning, C., Robinson, E., \& Gelino, D.\ 2013, AJ, 145, 21
\reference{} Kim, C., Kalogera, V., \& Lorimer, D.\ 2010, New Astr. Rev., 54, 148
\reference{} Kim, C., Perera, B., \& McLaughlin, M.\ 2013, MNRAS, submitted (arXiv:1308.4676)
\reference{} Kuulkers, E., et al., A\&A, submitted (arXiv:1204.5840)
\reference{} Lai, D.\ 2001, in Physics of Neutron Star Interiors, ed. by D. Blaschke, 
             N.K. Glendenning and A. Sedrakian, Lecture Notes in Physics, vol. 578, p.424 
\reference{} Lattimer, J., \& Prakash, M.\ 2010, To appear in Gerry Brown's 
             Festschrift; Editor: Sabine Lee (World Scientific)
             (arXiv:1012.3208)
\reference{} Li, Z., Qu, J., Song, L., Ding, G., \& Zhang, C.\ 2013, MNRAS, 428, 1704
\reference{} Macias, P., et al.\ 2011, in AAS 217, Vol. 43, 143.04
\reference{} Martin, R., Reis, R., \& Pringle, J.\ 2008, MNRAS, 391, L15
\reference{} Mennekens, N., \& Vanbeveren, D.\ 2013, A\&A, submitted (arXiv:1307.0959)
\reference{} Meurs, E, \& van den Heuvel, E.\ 1989, A\&A, 226, 88
\reference{} Mirabel, F., \& Rodrigues, I.\ 2003, Science, 300, 1119
\reference{} Nakar, E.\ 2010, proceedings of the The Shocking Universe meeting 
             (arXiv:1009.4648)
\reference{} Orosz, J., Jain, R., Bailyn, C., McClintock, J., \& Remillard,
             R.\ 1998, ApJ, 499, 375
\reference{} Orosz, J., McClintock, J., Aufdenberg, J., Remillard, R., Reid, M., 
             Narayan, R., \& Gou, L.\ 2011, ApJ, 742:84.
\reference{} O'Shaughnessy, R., Kim, C.\ 2010, ApJ, 715, 230
\reference{} Panter, B., Jimenez, R., Heavens, A., \& Charlot, S.\ 2008, 
             MNRAS, 391, 1117  
\reference{} Peters, P., \& Mathews, J.\ 1963, Phys. Rev., 131, 435
\reference{} Petrillo, C., Dietz, A., \& Cavagila, M.\ 2013, ApJ, 767, 140
\reference{} Reid, M., et al.\ 2011, ApJ, 742, 83
\reference{} Repetto, S., Davies, M., \& Sigurdsson, S.\ 2012, MNRAS, 425, 2799
\reference{} Repetto, S., Nelemans, G., in prep.
\reference{} Reynolds, M., Callanan, P., \& Filippenko, A.\ 2007, MNRAS, 374, 657
\reference{} Shahbaz, T.;, Ringwald, F., Bunn, J., Naylor, T., Charles, P., \& Casares,
             J.\ 1994, MNRAS, 271, L10  
\reference{} Steeghs, D., et al.\ 2013, ApJ, 768, 185
\reference{} Tauris, T., \& Takens, R.\ 1998, A\&A, 330, 1047
\reference{} Webbink, R.\ 1984, ApJ, 277, 355
\reference{} Weisberg, J., \& Taylor, J.\ 2005, in Binary Radio Pulsars, ASP 
             Conference Series, Vol. 328, Edited by F. A. Rasio and I. H. Stairs. 
             San Francisco: Astronomical Society of the Pacific, p.25 
\reference{} Wiktorowicz, G., Belczynski, K., Maccarone, T.\ 2013, ApJ, submitted (arXiv:1312.5924)
\reference{} Willems, B., Henninger, M., Levin, T., Ivanova, N., Kalogera, V., McGhee, K., Timmes, F.X.,
        \& Fryer, C.L.\ 2005, ApJ, 625, 324 
\reference{} Wong, T.-W., Valsecchi, F., Fragos, T., \& Kalogera, V.\ 2012, ApJ, 747, 111 
\reference{} Xu, X., \& Li, X.\ 2010, ApJ, 716, 114


\end{references}
\end{document}